\pdfoutput=1
\relax
\documentclass[letterpaper]{article} % DO NOT CHANGE THIS
\usepackage{aaai21}  % DO NOT CHANGE THIS
\usepackage{times}  % DO NOT CHANGE THIS
\usepackage{helvet} % DO NOT CHANGE THIS
\usepackage{courier}  % DO NOT CHANGE THIS
\usepackage[hyphens]{url}  % DO NOT CHANGE THIS
\usepackage{graphicx} % DO NOT CHANGE THIS
\usepackage{natbib}
\usepackage{soul}
\usepackage{caption} % DO NOT CHANGE THIS AND DO NOT ADD ANY OPTIONS TO IT
\usepackage{amsmath}
\usepackage{color}    
\usepackage{subfigure}
\usepackage{float}
\begin{document}
\title{{\LARGE Detecting Parkinson’s Disease From an Online Speech-task}}

\author{Wasifur Rahman$^a$\footnote{echowdh2@ur.rochester.edu}, Sangwu Lee$^a$, Md. Saiful Islam$^b$, Victor Nikhil Antony$^a$, Harshil Ratnu$^a$, \\Mohammad Rafayet Ali$^a$, Abdullah Al Mamun$^a$, Ellen Wagner$^c$, Stella Jensen-Roberts$^c$, \\Max A. Little$^{d,e}$, Ray Dorsey$^c$, and Ehsan Hoque$^a$
\\
\vspace{5mm}
{\normalfont
    $^a$ Department of Computer Science, University of Rochester, United States\\
    $^b$ Department of Computer Science and Engineering, BUET, Bangladesh\\
    $^c$ Center for Health + Technology, University of Rochester Medical Center, United States\\
    $^d$ School of Computer Science, University of Birmingham, United Kingdom\\
    $^e$ MIT Media Lab, MIT, United States
    }
}

\maketitle
%\email{echowdh2@ur.rochester.edu}
%\affiliation{Department of Computer Science, University of Rochester}
%\linenumbers
%alternative titles
%\title{Predicting Parkinson's in the Wild Using Speech and Continuous Vowel Phonation}
% We probably want to avoid "predicting" and rather opt for "characterizing"
% The VAANGO Visual Audio Analysis Network Gauge for Online Analysis of Parkinson's Disease Risk

% \author{Wasifur Rahman}
% \email{echowdh2@ur.rochester.edu}
% \affiliation{
%   \institution{University of Rochester}
%   \streetaddress{250 Hutchison Rd}
%   \city{Rochester}
%   \state{NY}
%   \postcode{14620}
% }
% \author{Jianyuan Zhong}
% \email{jzhong9@u.rochester.edu}
% \affiliation{
%   \institution{University of Rochester}
%   \streetaddress{250 Hutchison Rd}
%   \city{Rochester}
%   \state{NY}
%   \postcode{14620}
% }
% \author{Mohammad Rafayet Ali}
% \email{mali7@cs.rochester.edu}
% \affiliation{
%   \institution{University of Rochester}
%   \streetaddress{250 Hutchison Rd}
%   \city{Rochester}
%   \state{NY}
%   \postcode{14620}
% }
% \author{Taylan Sen}
% \email{tsen@cs.rochester.edu}
% \affiliation{
%   \institution{University of Rochester}
%   \streetaddress{250 Hutchison Rd}
%   \city{Rochester}
%   \state{NY}
%   \postcode{14620}
% }
% \author{Mohammed (Ehsan) Hoque}
% \email{mehoque@cs.rochester.edu}
% \affiliation{
%   \institution{University of Rochester}
%   \streetaddress{250 Hutchison Rd}
%   \city{Rochester}
%   \state{NY}
%   \postcode{14620}
% }
%\sloppy

\begin{abstract}

%Kurt --> Consider this alternative abstract: 
%Early detection of Parkinson's disease is crucial for improving the quality of life for those afflicted individuals. Unfortunately, studies have shown that between 25 and 80% of individuals with PD are fighting their battles alone as the disease remains undiagnosed and untreated. In this paper we show the promise of a web application specifically tailored to collect in the wild audio samples and assess an individual's risk for developing PD. We collect data from 726 unique participants (262 PD and 464 non-PD) uttering "the quick brown fox ... " to conduct an automated PD assessment. From the speech data, we extract both standard acoustic features and deep learning based embedding features to train several machine learning models. We engineer the algorithms of our models to generate an output that is both interpretable and explainable. This quality further enhances the value of our web application as a useful diagnostic tool for PD. 

 In this paper, we envision a web-based framework that can help anyone, anywhere around the world record a short speech task, and analyze the recorded data to screen for Parkinson's disease (PD). We collected data from 726 unique participants (262 PD, 38\% female; 464 non-PD, 65\% female; average age: 61) -- from all over the US and beyond. A small portion of the data was collected in a lab setting to compare quality. The participants were instructed to utter a popular pangram containing all the letters in the English alphabet ``the quick brown fox jumps over the lazy dog $\cdots$''. 
%  \st{Although the majority of subjects are above the age of 50, there is a healthy presence of younger subjects.}
 We extracted both standard acoustic features (Mel Frequency Cepstral Coefficients (MFCC), jitter and shimmer variants) and deep learning based features from the speech data. Using these features, we trained several machine learning algorithms. We achieved 0.75 AUC (Area Under The Curve) performance on determining presence of self-reported Parkinson's disease by modeling the standard acoustic features through the XGBoost -- a gradient-boosted decision tree model. Further analysis reveal that the widely used MFCC features and a subset of previously validated dysphonia features designed for detecting Parkinson's from verbal phonation task (pronouncing `ahh $\cdots$') contains the most distinct information. Our model performed equally well on data collected in controlled lab environment as well as `in the wild' across different gender and age groups. Using this tool, we can collect data from almost anyone anywhere with a video/audio enabled device, contributing to equity and access in neurological care.
 
%  In future work, we should seek to expand our geographic reach.

% Our model performed equally well on data collected in controlled lab environment as well as `in the wild' across different gender and age groups, potentially contributing to equity and access in neurological care. 

% \MSI{``We also provide explanation behind our model's decision and show that it is focusing mostly on the widely used MFCC features and a subset of dysphonia features previously used for detecting PD from verbal phonation task.'' -- is this important enough to go into Abstract? I have missed this point in Introduction. Can you add a few lines about it?}

% \MSI{We need to stress on the access part in the abstract, which is not present now. Our major contribution will be, our detection system is designed for the common people by making it easy to access and requiring a minimal supervision. I am a little doubtful about using AUC, since it may not be known in the medical domain, and abstract does not seem to be a good place to explain AUC.}

\end{abstract}

% <concept_id>10003120</concept_id>
% <concept_desc>Human-centered computing</concept_desc>
% <concept_id>10003120.10003121.10003122.10010855</concept_id>
% <concept_desc>Human-centered computing~Heuristic evaluations</concept_desc>
% <concept_significance>300</concept_significance>
% <concept_id>10003120.10003121.10003124.10010868</concept_id>
% <concept_desc>Human-centered computing~Web-based interaction</concept_desc>
% <concept_significance>400</concept_significance>
% <concept_id>10003120.10003138.10003139.10010904</concept_id>
% <concept_desc>Human-centered computing~Ubiquitous computing</concept_desc>
% <concept_id>10003120.10011738.10011776</concept_id>
% <concept_desc>Human-centered computing~Accessibility systems and tools</concept_desc>

% \ccsdesc[500]{Human-centered computing}
% \ccsdesc[500]{Human-centered computing~Ubiquitous computing}
% \ccsdesc[500]{Human-centered computing~Accessibility systems and tools}
% \ccsdesc[400]{Human-centered computing~Web-based interaction}
% \ccsdesc[300]{Human-centered computing~Heuristic evaluations}

% \keywords{Parkinson's disease, Datasets, Neural Networks, Machine Learning, Audio Classification, Audio Processing, Speech Recognition, Accessible, Remote, MDS-UPDRS}

\maketitle

\section{Introduction}
%\MSI{Why should we consider Parkinson's disease seriously?}\\
%\MSI{Why early and remote detection is necessary?} \\
%\MSI{Why do we need a detection model that can be accessed by all?} \\
%\MSI{What has been done in this direction? Why speech analysis is popular?}\\
% \W{Dr Hoque outline:
% PD affects more males than females. PD affects the aging population. As a result, it is easy for our algorithm to have more data for the prevalent condition leading to a framework that is not inclusive. Additionally, most of the training data is collected in a lab environment making it difficult for the deployed system to be useful outside of the lab. In this paper, we propose and validate a model that is trained on noisy real-world data and inclusive of all genders and age groups. To do this, we use a technique called SHAP because it’s the only feature attribution method that fulfills the mathematical definition of fairness. Our algorithm builds on prior work on speech in PD and our results provide support for prior findings. Our proposed unified model and the stratified gender-specific models have similar performance.  }

The Parkinson's disease (PD) is the fastest growing neurological disease in the world. Unfortunately, an estimated 20\% of PD patients remain undiagnosed. This can be largely attributed to the shortage of neurologists worldwide~\cite{khadilkar2013neurology,howlett2014neurology}, and limited access to the healthcare. An early diagnosis and continuous monitoring which allows for adjusting medication dosage are the keys to managing the symptoms of this incurable disease. The current standard of diagnosis requires in-person clinic visits where an expert assess the disease symptoms while observing the patients perform tasks from the Unified Parkinson's Disease Rating Scale (MDS-UPDRS)~\cite{goetz2008movement}. The MDS-UPDRS includes 24 motor related tasks to assess speech, facial expression, limb movements, walking, memory and cognitive abilities. Although many work has shown success by analyzing the hand-movements~\cite{ali2020spatio}, limb movement pattern~\cite{Lonini2018}, and facial expressions~\cite{bandini2017analysis}, speech is especially important because around 90\% of PD patients exhibit vocal impairment~\cite{ho1998speech, logemann1978frequency}, which is often one of the earliest indicators of PD~\cite{duffy2019motor}.

In this paper, we present our analysis of 726 audio recording of speech from 262 individuals with PD and 464 without. The speech recordings were collected using a web-based tool called \emph{Parkinson’s Analysis with Remote Kinetic-tasks} (PARK)\footnote{www.parktest.net}. The PARK tool instructed them to utter a popular pangram containing all the letters in the English alphabet, ``the quick brown fox jumps over the lazy dog...'' and recorded it. This allowed us to rapidly collect dataset that is more likely to contain the real-world variability associated with geographical boundaries, socio-economic status, age-groups and a wide variety of heterogeneous recording devices. The findings in this paper build on this unique real-world dataset and thus, we believe, could potentially generalize for real-world deployments.  

Collecting audio data from individuals often require in-person visits to the clinic limiting the number of data points as well as the diversity within the data. Recent advancement has allowed collecting tremor data from wearable sensors~\cite{kubota2016machine} as well as sleep data from RF radio signals~\cite{Yuebodycompass}. The existing work with speech and audio analysis utilizes sophisticated equipment for collecting data which are often noise free~\cite{tsanas2009accurate,little2008suitability} and do not contain the real-world variability. Given that a significant portion of the population has access to a mobile device with recording capability (for example, 81\% Americans own a smartphone~\cite{pew2019mobile}), we opted to use a framework allowing participants to record data from their home. From the recorded audio files, we have extracted acoustic features including Mel-frequency cepstral coefficients (MFCCs), known to represent the short-term power spectrum of a sound, jitter/shimmer variants (represents pathological voice quality), pitch related features, spectral power, and dysphonia related features, which are designed to capture PD-induced vocal impairment~\cite{little2008suitability}. Additionally, we extracted features from a deep-learning based encoder -- Problem Agnostic Speech Encoder (PASE)~\cite{Pascual2019} -- that represents the information contained in a raw audio instance through a list of encoded vectors.
These features are modeled with four different machine learning models  -- Support-Vector-Machine, Random Forest, LightGBM, and XGBoost -- to classify individuals with and without PD.

\begin{figure}
    % \centering
    \includegraphics[width=\linewidth]{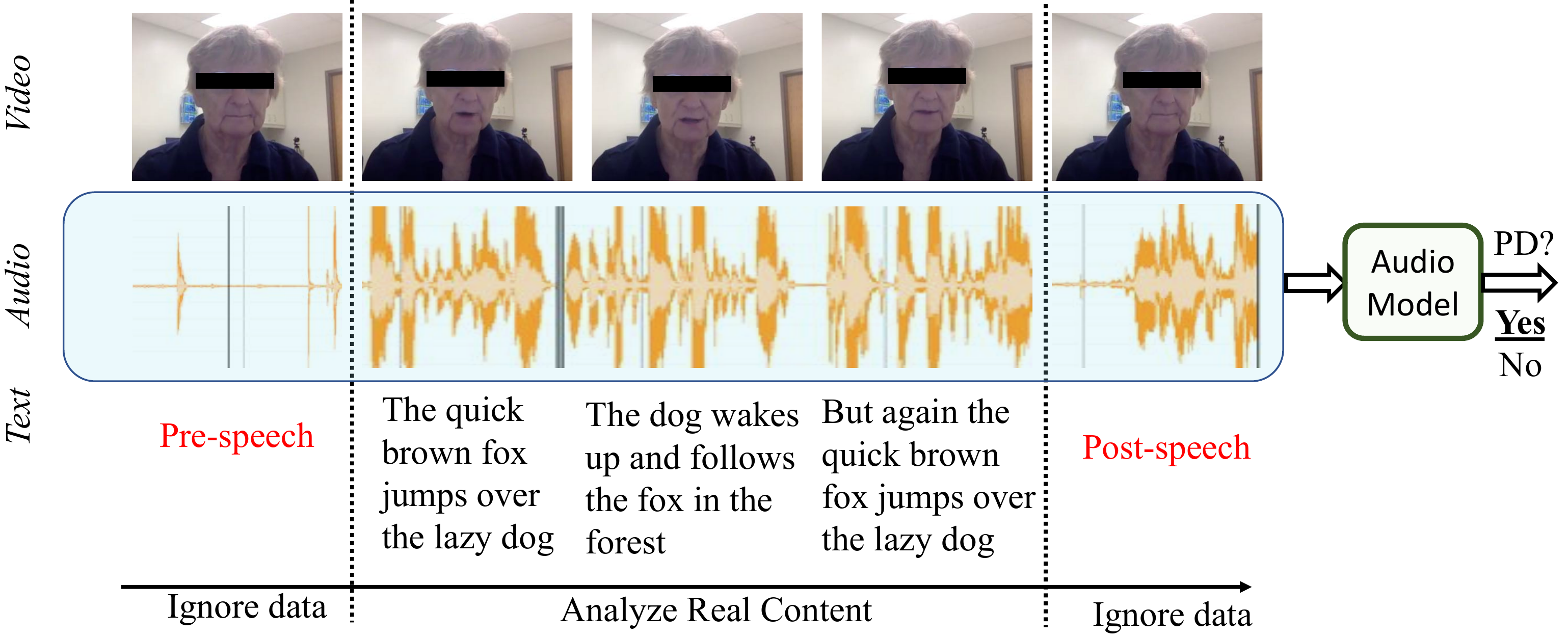}
\caption{An outline of our approach for solving the speech task of uttering ``The quick brown box $\cdots$''. While recording data, the participants often take some time to start recording the speech  and some more time to exit the data-collection system once the recording is done: those two segments are marked as pre-speech and post-speech respectively. We automatically remove these two components and analyze the \textbf{audio} data corresponding to completing the speech task to predict whether the subject has Parkinson's disease (PD).}
    \label{fig:teaser}
\end{figure}

% \MSI{Should we remove t=4s, t=12s -- these time labelling from Figure 1? These do not seem to add anything, rather may create confusion.}

% In this paper, we trained a model that can detect PD/non-PD from the recorded speech of our subjects pronouncing "the quick brown fox jumps over the lazy dog...”. Although automated PD detection can be done done by analyzing hand-movements~\cite{ali2020spatio}, movement pattern~\cite{Lonini2018}, or facial expressions~\cite{bandini2017analysis}; analyzing audio~\cite{little2008suitability,tsanas2012using,Arora2015,rueda2017feature} remains a promising avenue for PD detection. This is largely due to two important factors: around 90\% of PD patients exhibit vocal impairment~\cite{ho1998speech, logemann1978frequency} and this impairment is often one of the earliest indicators of PD~\cite{duffy2019motor}. 

Fig.\ref{fig:teaser} provides an outline of the data-analysis system. Our contributions can be summarized as follows:
\begin{itemize}
    \item We report findings from one of the largest dataset with real-world variability containing 726 unique participants mostly from their home. 
    
    \item We analyzed the audio features of speech to predict PD v.s. non-PD with 0.7533 AUC (Area Under The Curve) score.
    
    \item We provide evidence that our model prioritizes MFCC features and a subset of dysphonia features~\cite{little2008suitability,tsanas2012novel} consistent with prior literature. 
    
    \item Our model performs consistently well when tested on gender and age stratified data collected in controlled lab environment as well as `in the wild'.
    
\end{itemize}

\section{Results}

\begin{figure}
    % \centering
    \includegraphics[width=\linewidth]{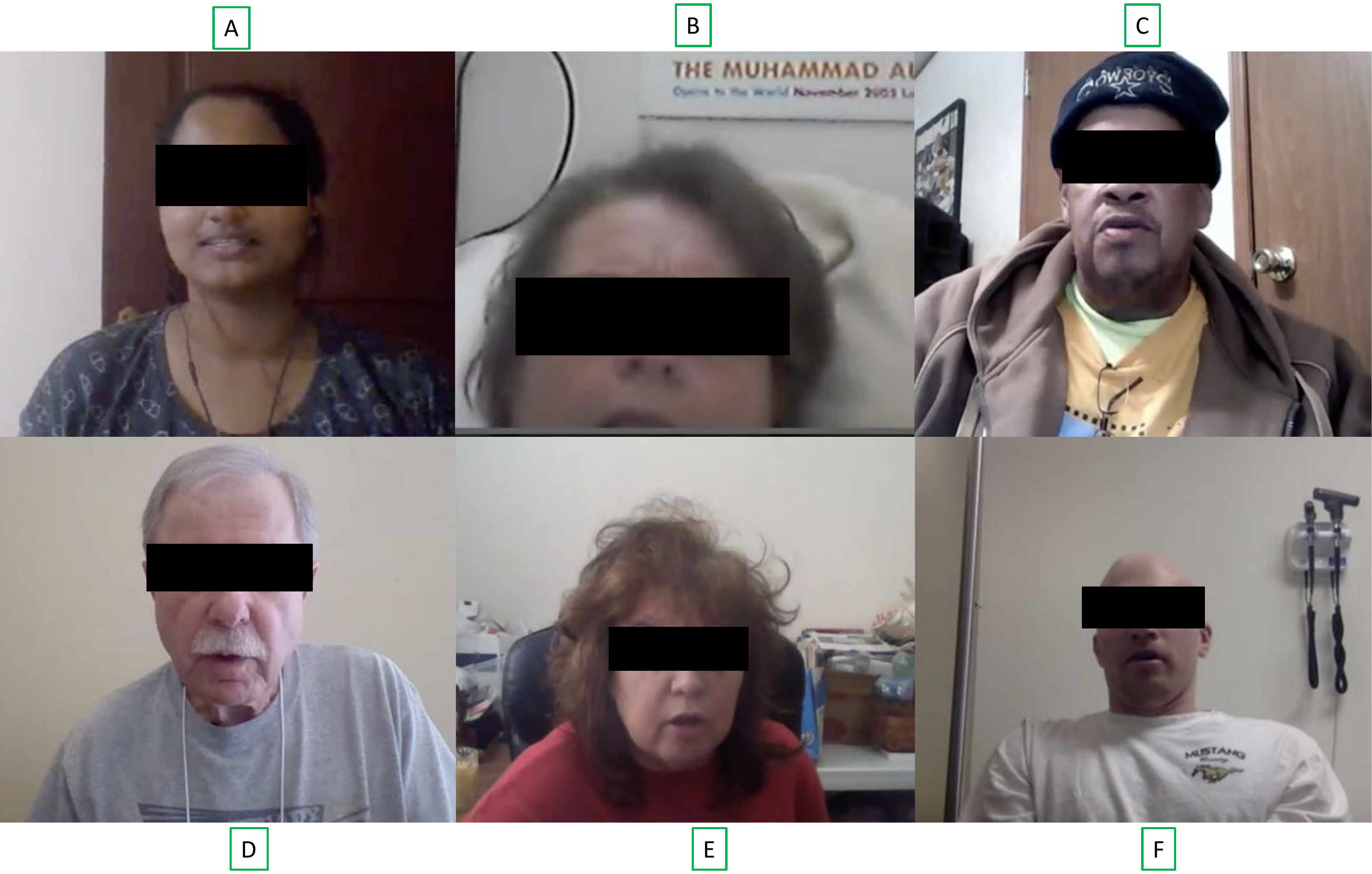}
\caption{Some screenshots of our subjects while providing the data. Subject A is from India, others are from the US. All the subjects except B provided data without any supervision. B, D, E, and F have been diagnosed with PD; B and F were diagnosed at the early age of 36 and 29 respectively. Electronic informed consent was taken from the participants to use their photos for publication.
% \W{Put A,B,C,,,. AND Provide cnew caption. saiful had some suggestions}From Left to Right, Top to Bottom: 20 year old female from India, not diagnosed with PD; 54 year old female from the US, diagnosed with PD at the age of 36; 59 year old male from the US, not diagnosed; 65 year old male from the US, diagnosed; 60 year old female from the US, diagnosed; 34 year old male from the US, diagnosed at the age of 29.
}
    \label{fig:data_samples_variety}
\end{figure}

We collected data from 726 \textit{unique} participants uttering the sentences \textit{``The quick brown fox jumps over the lazy dog. The dog wakes up and follows the fox into the forest, but again the quick brown fox jumps over the lazy dog''} using the Parkinson's Analysis with Remote Kinetic-tasks (PARK) \cite{langevin2019park} tool. Fig.~\ref{fig:data_samples_variety} contains some representative samples from our dataset. Table~\ref{tab:demography} shows the demographic information of the study participants. The gender distribution in the dataset is slightly skewed. Among all the participants, 55\% was female and 45\% was male. However, among participants with PD, only 38\% was female, and for non-PD, 65\% was female. Fig.~\ref{fig:age_pd_noon_pd_plot} shows the age distribution of the participants. We have a healthy balance between the number of PD/non-PD participants in the age range of [40-80] years, but most of the younger ([20-40] years) and older ([80-90] years) are from non-PD and PD groups respectively. Among the 726 participants, 54 completed the audio recording in a lab and the other 672 completed their recording at home. Having participants performing the tasks at home and at the lab allowed us to compare the results across both conditions. No participants appears in both sets and \textit{all} of our participants used the identical PARK protocol.

\begin{table}
  \centering
  \caption{Demographic Composition of our Dataset}
  \label{tab:demography}
  \resizebox{\columnwidth}{!}{
  \begin{tabular}{|l|l|l|}
    \hline
     & PD & non-PD\\\hline
    No. of participants ($N$) & 262 & 464\\\hline
    Female/Male & 101/161 & 300/164 \\\hline
    Age (mean $\pm$ std) & $65.92 \pm 9.2$ & $57.98 \pm 14.2$ \\\hline
    Country (US/other) & 199/63 & 419/45 \\\hline
    Years since diagnosed (mean $\pm$ std) & $7.88 \pm 5.41$ & N/A \\
    \hline
\end{tabular}
}
\end{table}

The data were pre-processed and both standard acoustic features (pitch, jitter, shimmer, MFCC, etc) and deep-learning based audio embedding features -- representing an audio clip as a feature vector -- were extracted; we will call these ~\textit{Standard-features} and \textit{Embedding-features} from now on. A complete list of ~\textit{Standard-features} is provided in Table ~\ref{tab:feature_names}, followed by a detailed description of the features in ~\ref{ssec:std_acoustic_feats}. We extract the \textit{Embedding-features} from PASE encoder~\cite{Pascual2019} that converts an audio signal into a representative vector. 

The rest of this section is organized as follows: results from the models built on entire dataset~\ref{ssec:res_entire_dataset}, interpretation of the best model (on entire dataset)~\ref{ssec:res_mod_inter}, and results and interpretation from specialized models on gender-stratified and age-matched datasets~\ref{ssec:gender_age_stratified_analysis}.

\subsection{Detecting Parkinson's Disease from Entire Dataset}
\label{ssec:res_entire_dataset}
To detect the PD patients from our dataset, we applied four machine learning algorithms: Support-vector-machine (SVM)~\cite{cortes1995support}, XGBoost ~\cite{chen2016xgboost}, LightGBM~\cite{ke2017lightgbm}, and Random-Forest~\cite{ho1995random}. We used a leave-one-out cross validation strategy where each data instance of the dataset is left out and the other data instances are used to create a model and predict the left-out instance iteratively. We used binary accuracy and Area-Under-Curve (AUC) metrics to report our model's performance. For a binary classifier, AUC denotes the area of the curve produced by plotting the true positive rate versus the false positive rate while varying the decision threshold of the model. Since the dataset is imbalanced, AUC is a better metric than accuracy to demonstrate the performance of our models. Table~\ref{tab:res_with_super_park} contains the AUC and accuracy score of the four machine learning models trained on the ~\textit{Standard-features} and ~\textit{Embedding-features} separately. Applying XGBoost on the Standard-features gave us the best performance of $0.75$ AUC and $0.74$ Accuracy. We also notice that models trained on interpretable \textit{Standard-features} work better than those trained on non-interpretable \textit{Embedding-features}.

\begin{table}
  \caption{\textit{Performance on the entire dataset:} The performance of various machine learning algorithms using  the Standard-features and Embedding-features on a dataset combining data from both Home-environment and Lab-environment. Models using Standard-features perform better than the models using  Embedding-features  in terms of both Binary Accuracy and AUC. Although the performance of the models are almost similar in terms of AUC metric, XGBoost outperforms others by considering both the AUC and Accuracy metrics simultaneously}
  \label{tab:res_with_super_park}
  \centering

\resizebox{\columnwidth}{!}{\begin{tabular}{|l|l|l|l|l|l|l|}
\hline
Algorithm &  \multicolumn{2}{|l|}{Standard-features} & \multicolumn{2}{|l|}{Embedding-features} \\ \hline
 &  AUC  & Accuracy & AUC  & Accuracy \\\hline
 
SVM & 0.751  & 0.735 & 0.738 & 0.692 \\
\hline
Random Forest  & 0.745  & 0.720 & 0.726  & 0.708 \\
\hline

LightGBM & 0.753  &
0.720 & 0.737  & 0.693 \\
\hline
XGBoost &  \textbf{0.750}  & \textbf{0.741} & 0.722  & 0.689 \\
\hline

\end{tabular}
}

\end{table}

% \begin{table*}
%   \caption{Results with SUPER PD}
%   \label{tab:praatResultsF1}
%   \centering
% \resizebox{\textwidth}{!}{
% \begin{tabular}{|l|l|l|l|l|l|l|l|l|l|l|l|l|l|l|}
% \hline
% Algorithm & \multicolumn{3}{|l|}{PRAAT} & \multicolumn{3}{|l|}{Max Little} & \multicolumn{3}{|l|}{Max Little + PRAAT} & \multicolumn{3}{|l|}{PASE} \\ \hline
%  & AUC & F1 & Acc. & AUC & F1 & Acc. & AUC & F1 & Acc. & AUC & F1 & Acc. \\\hline
% LightGBM & 0.6253 & 0.4715 & 0.6281 & 0.7599 & 0.5815 & 0.7163 & 0.7533 & 0.598 &
% 0.7204 & 0.7373 & 0.5654 & 0.6928 \\
% \hline
% XGBoost & 0.6209 & 0.4826 & 0.6185 & 0.7596 & 0.6074 & 0.7218 & 0.7502 & 0.6012 & 0.741 & 0.7219 & 0.5536 & 0.6887 \\
% \hline
% SVM & 0.6782 & 0.5312 & 0.6722 & 0.7542 & 0.6199 & 0.7314 & 0.7507 & 0.6175 & 0.7355 & 0.7379 & 0.6014 & 0.6915 \\
% \hline
% Random Forest & 0.6282 & 0.4824 & 0.6515 & 0.7513 & 0.5832 & 0.7438 & 0.745 & 0.5825 & 0.7204 & 0.7259 & 0.576 & 0.708 \\
% \hline
% \end{tabular}
% }
% \end{table*}

\subsection{Model Interpretation}
\label{ssec:res_mod_inter}
\begin{figure*}
    \centering
   
    \includegraphics[width=\textwidth]{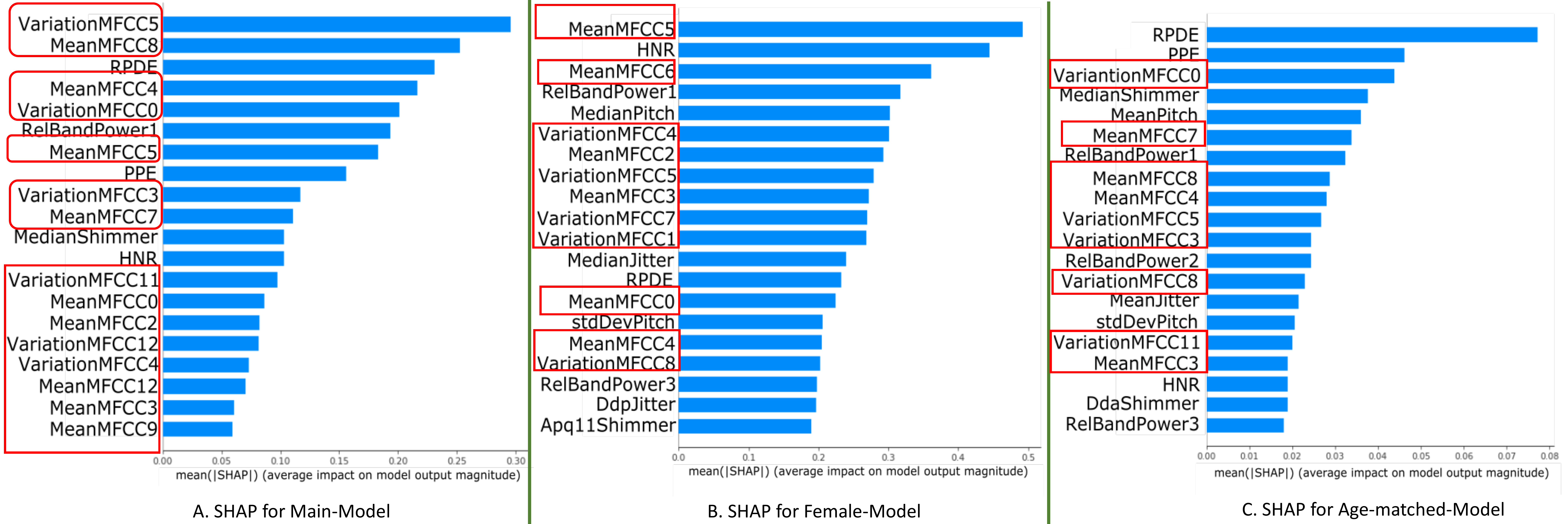}
    %  \subfigure[]{\includegraphics[width=\linewidth]{SHAP/MAX-PRAAT_Bar.png}} 
    %\caption{The three figures above depict SHAP analysis of our best performing model on three separate datasets:(A) All data present; (B) Female data only; (C) Age-matched dataset (all subjects are above age 50). The MFCC related features are put within a red box for increased visibility.  In \st{\textit{A.SHAP for Main-Model}}(A),  the most salient features are usually the MFCC features. Apart from that, features like RPDE (measuring uncertainty in F0 estimation), PPE (measure of inability of maintaining a constant F0), HNR (Harnomic to noise ratio), MedianShimmer (median in amplitude perturbation), RelBandPower1 (amount of power in the frequency range 0.5-1kHz) also impact the model's behaviour significantly. In \st{Part \textit{B.SHAP for Female model:}}(B), the most impactful features are the MFCC features as well. RPDE, HNR, MedianShimmer, relative-band-power features (RealBandPower1 and RelBandPower3: amount of power in the frequency range 0.5-1kHz and 2kHz-4kHz) also impact the model's behaviour. Besides, there are some influence from the pitch related features as well (MedianPitch, MedianJitter, StdDevPitch). In \st{\textit{C.SHAP for Age-matched model:}}(C), the model is greatly influenced by dysphonia features (RPDE and PPE), MFCC features and Pitch related features (MeanPitch, MeanJitter, StdDevPitch). HNR, MedianShimmer, relative-band-power features \st{(RelBandPower1 ,RelBandPower2 and RelBandPower3)} also impact the model's behaviour.  }
    
    \caption{We briefly describe the SHAP analysis of our best performing models on three datasets: (A) entire dataset, (B) female only, and (C) age matched (all subjects are above age 50). The MFCC features (put within a red box) are highly significant in all cases. Additionally, \texttt{RPDE} (measuring uncertainty in F0 estimation), \texttt{PPE} (measure of inability of maintaining a constant F0), \texttt{HNR} (Harnomic to noise ratio), \texttt{MedianShimmer} (median in amplitude perturbation), relative-band-power features (\texttt{RelBandPower1}, \texttt{RelBandPower2}, \texttt{RelBandPower3} indicating the amount of power in several frequency ranges) also impact the model's behaviour significantly. Pitch related features like \texttt{MedianPitch}, \texttt{MedianJitter}, \texttt{StdDevPitch} are also important for female and age matched models. Dysphonia features (\texttt{RPDE} and \texttt{PPE}) are the most important for the age matched model.
    }
    \label{fig:bar_three_models}
\end{figure*}

To further focus on the clinical implications of our work, we wanted to interpret the decisions of our classifiers. We use SHAP (SHapley Additive exPlanations)~\cite{lundberg2017unified,lundberg2020local} to recognize the features that are driving the model's performance. We choose SHAP for two reasons: it is well-suited for explaining the output of any machine learning model; it's the only feature attribution method that fulfills the mathematical definition of fairness. The goal of SHAP is to explain model's prediction of any instance as a sum of contributions from it's feature values; if a data-instance can be thought of as: $X_i = [f_1,f_2,\dots f_N]$, SHAP will assign a number to each of these $f_j$ features, denoting the impact of that feature -- both the magnitude and direction -- on the model's prediction. 
% \hl{An example of such a local explanation for an individual data instance is provided in Appendix. Sangwu: Remove since appendix is gone?}
Then all these local explanations are aggregated to create a global interpretation for the entire dataset. That global interpretation is presented in Fig.~\ref{fig:bar_three_models}.A, top 20 most impactful features -- ranked by having the most impact to the least -- are presented. To calculate each feature's impact, all of its SHAP values across all the data instances are gathered, and then the mean of their absolute values is calculated. A more technical description of SHAP is provided in Section \ref{ssec:methods_model_interpretation}.

Features that impacted the model's performance are typically the spectral features: the mean values or the variation of MFCC in each spectrum range. Apart from that, some other complex features such as, RPDE (measure of uncertainty in F0 estimation), PPE (measure of inability to maintain a constant F0), and HNR (Harmonic to Noise Ratio) also impacted the model's decision.

\subsection{Gender and Age Stratified Analysis}
\label{ssec:gender_age_stratified_analysis}
\begin{figure}
    % \centering
    \includegraphics[width=\linewidth]{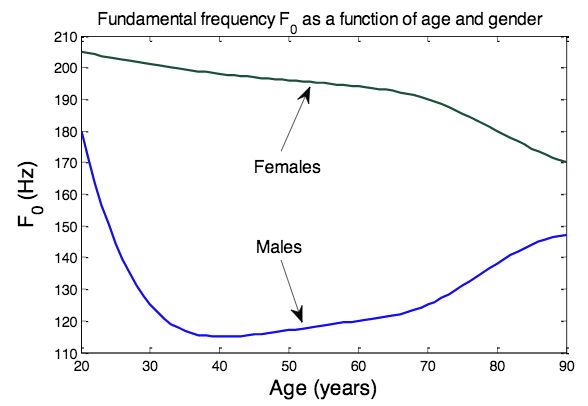}
    \caption{Changes in fundamental frequency F0 of voice as a function of gender and age  (collected from Tsanas et al.\cite{tsanas2009accurate}, who took the inspiration from Titze et al. \cite{titze1998principles} ). Female voices have higher F0 value but it decreases with age. Males typically have higher F0 value in their youth, with decreases with age and then increases roughly after the age of 45.}
    \label{fig:f0_by_age_gender}
\end{figure}

The characteristics of a person's voice is greatly influenced by their age and gender. In Fig.\ref{fig:f0_by_age_gender}, we see that males and females display a changing characteristics in their voice as they get older. Therefore, it can produce confounding effects in analyzing PD from audio where the machine learning model uses audio features to detect PD. To minimize the effect of confounding factors, researchers in the past trained separate models on data from male and female participants~\cite{tsanas2012using} or analyzed an age-matched dataset by considering data from participants above the age of 50~\cite{ali2020spatio,langevin2019park}.

\subsubsection{Building specialized models for each gender and age-matched analysis}
\begin{table*}[]
\centering
\caption{Gender and age stratified models: Three separate datasets are constructed: a Male dataset with male subjects, a Female dataset with female subjects, and age-matched dataset by excluding the subjects below the age of 50. For each of these datasets, a separate model is constructed and its performance reported below.}
  \label{tab:gender_age_stratified_models}
\begin{tabular}{|l|l|l|l|l|l|l|}
\hline
Algorithm     & \multicolumn{2}{l|}{Male} & \multicolumn{2}{l|}{Female} & \multicolumn{2}{l|}{Age-matched} \\ \hline
              & AUC        & Accuracy     & AUC         & Accuracy      & AUC        & Accuracy    \\ \hline
SVM           & \textbf{0.795}      & \textbf{0.717}       & 0.659      & 0.763        & \textbf{0.755}     & \textbf{0.723}      \\ \hline
Random Forest & 0.758     & 0.702       & 0.699      & 0.788         & 0.739     & 0.713      \\ \hline
LightGBM      & 0.725     & 0.665       & \textbf{0.717}      & \textbf{0.768}        & 0.7498     & 0.7116      \\ \hline
XGBoost       & 0.762     & 0.717       & 0.682      & 0.771        & 0.742     & 0.704      \\ \hline
\end{tabular}
\end{table*}
The performance metrics of the machine learning models trained on male, female and age-matched dataset are in Table~\ref{tab:gender_age_stratified_models}. By comparing the performance with metrics presented in Table~\ref{tab:res_with_super_park}, we can see that the models that used male or age-matched datasets performed in-par or better than the models that used the whole dataset to train. However, there is a performance drop in the models using female dataset. Table~\ref{tab:demography} shows that females are over-represented in the non-PD group and under-represented in the PD group, leading to data-imbalance and possibly lowering performance for the female-only model. 

% The shap weight got from the female model only.
% \input{sections/figure_latex/female_model_shap_weight}

We also analyzed the features that are driving these specialized models' performance through SHAP analysis. Fig \ref{fig:bar_three_models}.B displays the most salient features ranked by their SHAP value and the distribution of how each feature impacts the model's decision making. The most important features are still dominated by the MFCC related features or complex features like HNR (Harmomic-to-Noise Ratio), relative-band-power in different frequency ranges ( RelBandPower1, RelBandPower3), RPDE (uncertainty in F0 estimation), perturbation in F0 (DdpJitter) or Perturbation in amplitude (Apq11Shimmer). However, one noticeable fact is that three pitch and jitter related features: MedianPitch (median principal frequency, StdDevPitch (Standard-deviation in principal frequency) and MedianJitter (median variation in F0) are also impacting the model's prediction which was not noticed in the SHAP analysis run on the All-data-model.

% According to SHAP, higher value of median frequency (f0m), higher median variation in F0 (f0jr), and lower standard-deviation in F0 (f0std) will push the model towards predicting someone as a PD patient.

Similarly, we interpreted the salient features for the Age-matched dataset in Fig ~\ref{fig:bar_three_models}.C. We noticed that the most salient features are usually coming from the MFCC feature-groups, complex features( RPDE, PPE, HNR), relative-band-power (RelBandPower1, RelBandPower2, RelBandPower3), and pitch related features. We also see that the pitch related features are also driving the model's prediction as well.

\section{Discussion and Future Work}

\subsection{Detecting PD From Regular Conversation}

Some of the most common voice disorders induced by PD are:
dysphonia (distortion or abnormality of voice), dysarthria (problems with speech articulation), and hypophonia (reduced voice volume). Two speech-related diagnostic tasks are commonly used for detecting PD through exploiting the changing vocal pattern caused by these disorders: (i) sustained phonation (the subject is supposed to utter a single vowel for a long time with constant pitch), and (ii) running speech (the subject speaks a standard sentence). Little et al.~\cite{little2008suitability} developed features for detecting dysphonia from people with PD. Tsanas et al.~\cite{tsanas2009accurate} focused on the telemonitoring of self-administered sustained vowel phonation task to predict the UPDRS rating~\cite{movement2003unified} -- a commonly used indicator for quantifying PD symptoms. These studies train their models with data captured by sophisticated devices (e.g., wearable devices, high resolution video recorder, Intel at-home-testing-device telemonitoring system) that are often not accessible to all and difficult to scale. The performance of these models can reduce significantly when classifying data collected in home acoustics. Additionally, completing the sustained phonation task correctly requires following a specific set of guidelines such as completing the task in one breath -- which can be difficult for older individuals.

In contrast, we analyze the running speech task from the data collected by using a web-based data collection platform that can be accessed by anyone, anywhere in the world and requires only an internet connected device with integrated camera and microphone. Besides, running speech task does not require conforming to specific instructions, and are more similar to the regular conversation, and therefore, the model can be potentially augmented to predict PD from regular conversation -- a potential game changer in PD assessment. In the future, user-consented plug-ins could be developed for any application such as Alexa, Google Home, or Zoom where audio is transmitted between persons. Anyone who consents to download the plug-in and uses it while they are on the phone, over zoom, or giving virtual/in-person presentations could benefit from receiving an informal referral to see a neurologist, when appropriate. 

% \W{May be added here somewhere:Due to collecting data from home environment, the data is very noisy and posits a lot of different challenges: lower audio quality, incomplete compliance with the the instructions, background noise, variety in recording devices, etc.}
%%%%%%%%%%%%%%%%%%%%%%%%%%%%%%%%%%%%%%%%%%%%%%%%%%%%%%%%%%%5
\subsection{Validating Model Interpretation}
The features that SHAP found to be having impact of modeling decisions are well-supported by previous research. For example, MFCC features have already proven to be useful in a wide range of audio tasks such as speaker recognition~\cite{bhattarai2017experiments}, music information retrieval~\cite{muller2007information}, voice activity detection~\cite{kinnunen2007voice}, and most importantly, in voice quality assessment~\cite{tsanas2011nonlinear}. Similarly, The high impact of HNR (Harmonic to Noise ratio), RPDE (measuring uncertainty in F0 estimation), and PPE (measure of inability of maintaining a constant F0) on the model's output is in congruence with the findings from Little et al.~\cite{little2008suitability}. However, explaining Fig.~\ref{fig:bar_three_models}(A) in the light of the PD-induced vocal impairment is a difficult task. MFCC features are calculated by converting the audio signal into the frequency domain; they denote how energy in the signal is distributed within the various ranges of frequency. Therefore, giving a physical interpretation to the SHAP values corresponding to the MFCC features is not straight-forward. Similarly, Little et al.~\cite{little2008suitability} designed the RPDE and PPE features for modeling the sustained phonation task (uttering `ahh $\cdots$') with the assumption that the healthy participants will be able to maintain a smooth and regular voice pattern. In contrast, uttering multiple sentences introduces a lot of variation in the data, adding a wide set of heterogeneous patterns. Therefore, the underlying assumptions behind constructing those features do not hold for our task of uttering multiple sentences.

\begin{figure}
    % \centering
    \includegraphics[width=\linewidth]{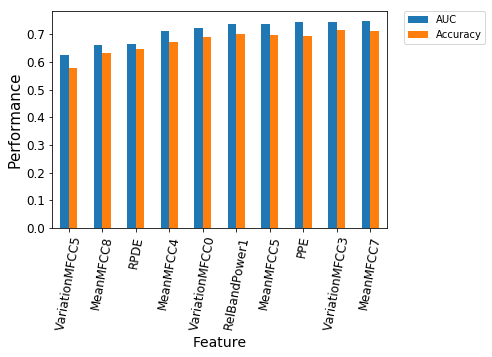}
\caption{Validating the SHAP output of Fig. ~\ref{fig:bar_three_models}.A: We started with the most important feature depicted in Fig.\ref{fig:bar_three_models} (VariationMFCC5) and added the next salient feature one at a time. For each feature set, we trained a new model and showed that model's performance in terms of AUC and accuracy. As evident from the Figure, after adding 8 most salient features (up to PPE), the primary metric, AUC's value is saturated.}  %\WS{Please give names to the axes}}
    \label{fig:validating_shap_itself}
\end{figure}

In Fig.\ref{fig:validating_shap_itself}, we present an empirical validation of the SHAP output presented in Fig.~\ref{fig:bar_three_models}.A. We incrementally add one feature at a time to build a dynamic feature-set, train successive models on that feature-set, and report the Accuracy and AUC performances. We can see that the performance of our model saturates after adding 7-8 features. Therefore, we can say that the SHAP analysis teases out the most important features driving the model's performance successfully.
% \W{CORRECT IT}
% A person without any voice impairments should have high HNR, low RPDE and low PPE in the sustained phonation task (e.g., pronouncing `Ahhh') for data collected through sophisticated devices like Intel AHTD~\cite{little2008suitability}. However, Little et al.~\cite{little2008suitability} also demonstrated that there is often significant overlap in probability distribution for each of these features between the PD and non-PD voice data. Similarly, if we look at Fig.\ref{fig:shap_three_models}.A, there is a lot of overlap on SHAP values for these three features exemplified by the mixed colors around SHAP values close to zero (either positive or negative). Besides, according to our SHAP analysis, the model's output value is increased by: lower values of RPDE, lower values of PPE and higher values of HNR. These findings are incongruent with those of ~\cite{little2008suitability}. We attribute this incongruence to background noise in our data, and difference between our speech task and previously researched sustained phonation task. However, we can still see that those features are important in determining PD from speech tasks. Therefore, further research is necessary to augment those features to work well in the running speech task. 

\subsection{Why Not Stratified Analysis Only?}
We built models inclusive of all genders for several reasons. First, there are potential shared characteristics among vocal patterns of both gender that can be relevant for detecting PD. Second, dividing the dataset into two portions will reduce available training data for each model, which may in turn reduce the generalization capability of each model. Besides, our model analyzes data from patients of all ages. Although most people who are diagnosed with PD are over the age of 60, about 10\%-20\% of the diagnosed PD population are under the age of 50, and about half of them are under the age of 40~\cite{earlypd}. As an anecdotal evidence, Michel J Fox was diagnosed with PD at the age of 29~\footnote{https://www.michaeljfox.org/michaels-story}, Muhammad Ali had PD by 42\footnote{https://parkinsonsnewstoday.com/2016/06/10/muhammad-alis-advocacy-parkinsons-disease-endures-boxing-legacy/}. This is unfortunate because these people have the longest to live with the PD symptoms. 
% Through a focused group discussion, we have discovered that most of the members of our research group know people who had PD below 40. 
In our dataset, there is also a minority of PD patients below the age of 50 (Fig.\ref{fig:age_pd_noon_pd_plot}). Based on these observations, we believe that our system should provide access to all people irrespective of age. PD does not discriminate by age while impacting a person, and an automated system should not discriminate based on age and should provide equitable service to people of all ages. However, these factors can work as confounders in PD analysis. Therefore, we provide additional analysis to ensure that our model is not using the idiosyncrasies of group-specific information to make prediction in \ref{ssec:gender_age_stratified_analysis}. 

\subsection{Performance excluding lab-environment data}
\begin{figure}
    % \centering
    \includegraphics[width=\linewidth]{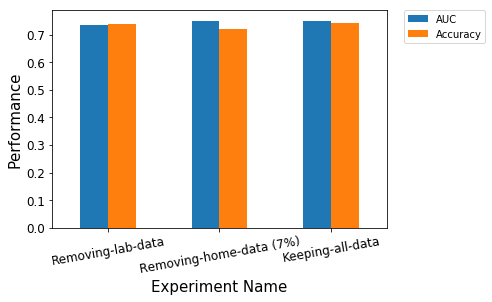}
\caption{The performance of our models are fairly consistent across three different experiments: Removing-lab-data, Removing-home-data (7\%), and Keeping-all-data.}  
    \label{fig:removing_lab_home_seven_percent}
\end{figure}

When the data was collected in the lab, the participants had access to a clinician providing support using consistent recording set up and dedicated bandwidth. On that contrary, the data collected in the home setting assumed no assistance, and included the real-world variability of heterogeneous recording set up and inconsistent internet speed. Theoretically, the data collected at the "lab" and "home" were very different from each other. 

To ensure that our model works equally well without the "clean lab data", we designed two experiments. In experiment 1, we removed clean-lab-data which is around 7\% of the entire dataset (54 data points), retrained our model on rest of the 672 participants with the leave-one-out validation procedure, and calculate the performance metrics. In experiment 2, we randomly remove 7\% (roughly 54 data-points) "home" data from the entire dataset (while keeping the "lab" data intact), building a model with the rest of 93\% data with the leave-one-out cross-validation method. Then, we report the \emph{average} performance from these 10 runs. Fig. \ref{fig:removing_lab_home_seven_percent} contains all these performance metrics, including the one achieved by keeping-all-data (Table \ref{tab:res_with_super_park}). Fig. \ref{fig:removing_lab_home_seven_percent} shows that the AUC metric across these three experiments is fairly consistent, with a very small 0.015 drop in AUC for removing-lab-data, demonstrating that our framework performs equally well across "lab" and "home" data.

\subsection{Label Inconsistency and Predicting Tremor Score}
Using the PARK~\cite{langevin2019park} protocol, we have collected one of the largest dataset of participants conducting a series of motor, facial expression and speech tasks following the MDS-UPDRS PD assessment protocol \cite{goetz2008movement}. Although we analyze only the speech task in this paper, the dataset can be potentially used to automate the assessment of a large set of MDS-UPDRS tasks and facilitate early stage PD detection; thus, improving the quality of life for millions of worldwide. However, deploying the data-collection protocol on the web and facilitating access for anyone, anywhere around the world comes at a cost. So far, all of our PD participants have been clinically verified to be diagnosed with PD (more in \ref{ssec:data_collection}). Therefore, the label of PD data-points are reliable. However, our non-PD participants have not gone through any clinical verification. Our data-collection protocol asks them appropriate questions to check whether they have been diagnosed with PD, and collects data when they answer in the negative. However, we can not discount the possibility of a small subset of our non-PD population being in the very early stage of PD, and are oblivious about it. At present, there is estimated to be around 1 million PD patients in the US, out of a population of 330 million~\cite{pdIncidenceStat} -- yielding a PD prevalence rate of 0.3\%. However, as our non-PD dataset is largely tilted towards people over age of 50, the rate in our dataset could be higher than 0.3\%. 
% However, since the overall prevalence of PD is low (less than 1\%), the number of individuals with undiagnosed Parkinson's in our control population is likely low.
Even if we consider a liberal 1\% prevalence rate, the number of individuals with undiagnosed Parkinson's in our control population is likely low (at most 4.6 persons). Therefore, we believe the non-PD data label to be generally reliable.
In future, we plan to model tremor score in [0-4] range for each task -- 0 for no tremor and 4 for severe tremor -- instead of binary label following the MDS-UPDRS protocol to address this problem more thoroughly.

\subsection{Building a More Representative Dataset}
The PARK protocol is web-enabled, allowing anyone with access to internet to contribute data. We plan to augment our dataset by adding more non-native English speakers, more females, and more PD participants. As our PD data is collected through contacts from local PD clinics and non-PD data through Amazon Mechanical Turk, majority of our participants are from the US or other English speaking countries. To make our model more robust on data from non-native English speakers, we are in process of collecting both PD and non-PD data from non-native English speaking countries. 

Our best model for female-data performs worse than its male counterparts -- as demonstrated in Table ~\ref{tab:gender_age_stratified_models}. We attribute this degraded performance PD/non-PD imbalance for female participants in our dataset: the PD/non-PD ratio for female is $101/300$ (Table.\ref{tab:demography}). Previous epidemiological studies have shown that both incidence and prevalence of PD are 1.5–2 times higher in men than in women~\cite{van2003incidence,haaxma2007gender}. Therefore, any randomly sampled dataset for PD will have more prevalence of male, contributing to models more biased towards male. Our immediate plan is to prioritize collecting balanced data from all genders, age and race across the geographical boundaries, leading to a balanced dataset. 

Our dataset also suffers from the ubiquitous problem of data imbalance in diagnostic tests: number of data from non-PD participants is 1.8 times more than their PD counterparts. Therefore, there is a risk that the model will be biased towards predicting the majority non-PD class as a default and yield a high False-Negative score. To address this, we plan to recruit more PD participants in future to make our dataset more balanced. 

\subsection{Increasing Model Performance}
Although we consider AUC to be a better metric for our dataset, our model performs 10\% better than always choosing non-PD as the prediction in terms of Binary-accuracy. To be practically deployable in clinical settings, the performance needs to be improved further. We will focus on four promising avenues: making the dataset balanced; designing better features capable of modeling the nuanced pattern in our data; making our model resilient to noise present in our data; and de-confounding the PD prediction from age and gender variables.

For removing noise, we plan to augment the techniques proposed in Poorjam et. al \cite{poorjam2019automatic} to automatically enhance our data quality by detecting the segments of data that conform to our experimental design. Besides, as discussed in \ref{ssec:gender_age_stratified_analysis}, gender and age can appear as confounding variables in PD prediction task. In this paper, we have showed that our unified model and the stratified gender-specific models have similar performance. However, we plan to build better models to systematically deconfound the effects of both the age and gender variables while benefiting from them simultaneously. We can do this by incorporating the causal bootstrapping technique -- a re-sampling method that takes into account the causal relationship between variables and negate the effect of spurious, indirect interactions -- outlined in \cite{little2019causal}.

\section{Methods}
\begin{figure}
    % \centering
    \includegraphics[width=\linewidth]{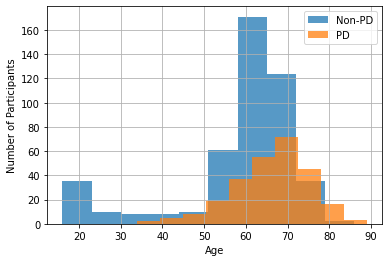}
    \caption{A bar plot showing age-distribution of  PD and non-PD subjects in our dataset. Although the number of non-PD subjects is 1.8 times the number of PD subjects, there is a healthy balance between these two groups in the age-range of [40-80] years. However, non-PD subjects outnumber their PD counterparts in the age group [20-40]. Similarly, PD subjects outnumber non-PD subjects significantly in the age group of [80-90] years.}
    \label{fig:age_pd_noon_pd_plot}
\end{figure}

%\MSI{Should we show the plot (Figure \ref{fig:age_pd_noon_pd_plot}) in a normalized way (y-axis showing probability instead of count)? We might be interested in presenting that the age distribution is similar for both categories.}
\begin{table*}
  \caption{Names of all the features, feature collection protocol and short-description of the features used by us.  The correlated features were removed and features with \textbf{bold} text  were used in building the models. Feature names are preceded by the loosely defined umbrella category they belong to.}
  \label{tab:feature_names}
  \centering
\resizebox{\textwidth}{!}{
\begin{tabular}{|l|l|l|l|l|l|l|}
\hline
Feature & Code-source & Short-description \\ \hline

Pitch:\textbf{MedianPitch} & ~\cite{little2007exploiting} & Median principal frequency  \\
\hline
Pitch:\textbf{MeanPitch} & ~\cite{praat} & Mean principal frequency \\
\hline
Pitch:\bf StdDevPitch & ~\cite{little2007exploiting} &Standard deviation in principal frequency\\ 
\hline
Jitter:\bf MeanJitter & ~\cite{little2007exploiting} & Perturbation in principal frequency (mean variation) \\
\hline
Jitter: \bf MedianJitter & ~\cite{little2007exploiting} & Perturbation in principal frequency (median variation) \\
\hline
Jitter:LocalJitter & ~\cite{praat} & Jitter variant\\
\hline

Jitter: \bf LocalAbsoluteJitter & ~\cite{praat} & Jitter variant \\
\hline

Jitter:RapJitter & ~\cite{praat} & Jitter variant \\
\hline

Jitter:Ppq5Jitter & ~\cite{praat} & Jitter variant \\
\hline

Jitter:\bf DdpJitter & ~\cite{praat} & Jitter variant \\
\hline
Shimmer:MeanShimmer & ~\cite{little2007exploiting} & Amplitude perturbation (using mean) \\
\hline
Shimmer: \bf MedianShimmer & ~\cite{little2007exploiting} & Amplitude perturbation (using median) \\
\hline
Shimmer:LocalShimmer & ~\cite{praat} & Shimmer variant \\
\hline

Shimmer:LocaldbShimmer & ~\cite{praat} & Shimmer variant \\
\hline

Shimmer:Apq3Shimmer & ~\cite{praat} & Shimmer variant \\
\hline

Shimmer:Apq5Shimmer & ~\cite{praat} & Shimmer variant \\
\hline

Shimmer: \bf Apq11Shimmer & ~\cite{praat} & Shimmer variant \\
\hline

Shimmer: \bf DdaShimmer & ~\cite{praat} & Shimmer variant \\
\hline

MFCC: \bf MeanMFCC[0-12] & ~\cite{little2007exploiting} & 13 features of mean MFCC \\
\hline
MFCC: \bf VariationMFCC[0-12] & ~\cite{little2007exploiting} & 13 features of mean variation of  MFCC \\
\hline
\bf RelBandPower[0-3] & ~\cite{tsanas2012novel} & Four features capturing relative band power in four spectrum ranges\\
\hline
\bf Harmonic to Noise ratio(HNR) & ~\cite{praat} & Signal-to-noise ratio \\
\hline
\bf Recurrence period density entropy(RPDE) & ~\cite{little2007exploiting} & Pitch estimation uncertainty \\
\hline
\bf Detrended Fluctuation Analysis(DFA) & ~\cite{little2007exploiting} & Measure of stochastic self-similarity in turbulent noise \\
\hline
\bf Pitch period entropy (PPE) &~\cite{little2008suitability} & Measure of inability of maintaining  constant pitch  \\
\hline
\end{tabular}
}
\end{table*}

\subsection{Data Collection }
\label{ssec:data_collection}
Our dataset is collected using the Parkinson's Analysis with Remote Kinetic-tasks (PARK) \cite{langevin2019park}  -- a web-based tool that guides user to conduct a series of motor, facial expression and speech tasks following the MDS-UPDRS PD assessment protocol \cite{goetz2008movement}. The users are recorded via webcam and microphone connected to the PC/Laptop and uploaded to a server while performing the tasks. In this work, we focus on the running speech task, where the participants were instructed to read a sentences -- \textit{"The quick brown fox jumps over the lazy dog. The dog wakes up and follows the fox into the forest, but again the quick brown fox jumps over the lazy dog"}. The first sentence is a pangram: it contains all the letters of the English alphabet; thus we get features relevant for pronouncing all phonemes for each subject. 

% By extending the PARK protocol, a new protocol named Super-PARK was developed. The participants, both PD patients and non-PD patients, visited a local clinic where they were instructed by a trained doctor/technician to conduct all the a MDS-UPDRS tasks. Although the data is collected using the same web interface, an external camera is used for recording participants. Due to in-person instruction and controlled data-collection environment, the Super-PARK data quality can be perceived as better than its PARK counterpart. We have built our primary models by incorporating data from both PARK and Super-PARK. However, to make sure that our model can perform well on the data collected in the home environment of our participants, we have conducted two additional sets of analysis by removing the Super-PARK data from the test-data and then from both training and test-data altogether. 
Table \ref{tab:demography} describes the demography of our participants in a nutshell. Figure \ref{fig:age_pd_noon_pd_plot} shows the age distributions of the participants. Among our 726 unique participants, 262 are diagnosed as PD patients; the rest are non-PD patients. We got the contact information of the PD patients from local clinics and various PD support groups. The non-PD participants are recruited from Amazon Mechanical Turk. During data-collection, informed consent of all the participants were taken. Among the 726 unique participants, only 54 gave data in the lab under the guidance of a study coordinator using the PARK tool; the rest of the 672 participants used the PARK system from their home to provide data. All the steps of the research -- subject recruitment, data collection, data storage and analysis -- were completed in accordance with the protocol agreed upon between the researchers and the Institutional Review Board (IRB) of the University of Rochester.

\subsection{Data Pre-processing}
\label{ssec:data_processing}

% These silent segments are trimmed using open-source pydub library. An audio segment is considered silent if its volume is under -40 dBFS for each 10ms interval. Only trimmed audio files with duration longer than 5 seconds are added to our dataset.

During data collection, the participants often took some time to start uttering the task sentences. After uttering the sentence, they often took additional time before stopping the recording. Hence, we have a substantial amount of noisy and irrelevant data at the beginning and at the end of most of the data instances (see Fig.\ref{fig:teaser}). To  remove those irrelevant data, we use Penn Phonetics Lab Forced Aligner Toolkit (P2FA)~\footnote{https://github.com/jaekookang/p2fa\_py3} toolkit. Given a audio file and transcript, it tries to predict the time boundaries where each of the words in the transcript was pronounced. If it cannot recognize a set of words, it skips them and outputs time boundaries for the words it could recognize successfully. The toolkit is built on the research done in ~\cite{yuan2008speaker}, where they apply a combination of Hidden Markov Models (from The Hidden Markov Model Toolkit (HTK) \footnote{http://htk.eng.cam.ac.uk/} and Gaussian-Mixture-Models(GMM) to align the words with the audio. The processing is done in several stages: HMM based models output the most likely sequence of hidden states for a given audio; Those hidden states are combined into phonemes and those phonemes are combined into words using a predefined dictionary (comprised of words and their corresponding phoneme pattern) through GMM based models. 

From the output of this system, we can get the starting time of first word that was recognized by the P2FA and the ending time of the last word recognized by P2FA. We use the audio segment in between them for further analysis. For building models capable of predicting PD, we extract two different sets of features: Acoustic-features (\ref{ssec:std_acoustic_feats}) including MFCC, Jitter variants, Shimmer variants, etc; and Embedding-features (\ref{ssec:pase_feats}) to represent an acoustic signal as a learned feature vector. The subsequent sections detail feature extraction process.

\subsection{Acoustic-features Extraction}
\label{ssec:std_acoustic_feats}

We extracted features by combining a subset of features collected through several sources: PRAAT features~\cite{praat} through the Parselmouth~\cite{parselmouth} python interface and the previously used features relevant for PD analysis in ~\cite{little2007exploiting,little2008suitability,tsanas2012novel}. After calculating all the features, we constructed a correlation  matrix of the feature values to calculate the degree of correlation between the features. Then, we iterated over each pair of features in an un-ordered fashion, and if the correlation co-efficient between them was over .9, we dropped one of those features from further analysis~\cite{goldberger1991course}. Table \ref{tab:feature_names} contains a short overview of the features used in our analysis; the feature names with bold text are the ones  used for building the models.
We provide a more comprehensive description of the features in the following sections. Some of our definitions are adapted from the official PRAAT documentation~\footnote{https://www.fon.hum.uva.nl/praat/}.

\subsubsection{Pitch Related Features:}
Pitch denotes the rate of vibrations present in a sound.
\emph{MedianPitch} and \emph{MeanPitch} denote the median and mean fundamental frequency or pitch of the audio signal. \emph{StdDevPitch} denotes the standard deviation of fundamental frequency f0.

\subsubsection{Jitter related features:}
Jitter defines how much a signal deviates from its presumed true periodicity; it is often an undesired quantity if our signals are assumed to be periodic.
\emph{MeanJitter} is the measure of jitter collected by calculating the mean variation of f0.
\emph{MedianJitter} is the jitter measure calculated using the median variation of f0. \emph{LocalJitter} denotes the average of the absolute differences between consecutive period of a signal -- divided by the average period. \emph{RapJitter} --  Relative Average Perturbation--  is computed by the average absolute difference between a period and the average of that period and the two neighbouring periods; divided by the average period. \emph{Ppq5Jitter} denotes the five-point Period Perturbation Quotient: the average absolute difference between a period and the average of it and its four closest neighbours -- divided by the average period. \emph{DdpJitter} denotes the average absolute difference between consecutive differences between consecutive periods, divided by the average period. 

\subsubsection{Shimmer Related Features:}
Shimmer is a measurement of amplitude instability in an audio signal; a normal voice will have minimal instability during sustained verbal phonation production. \emph{MeanShimmer} is the Shimmer value by quantifying the mean variation of amplitude in voice signals. \emph{MedianShimmer} is the Shimmer calculated using the median variation of amplitude. \emph{LocalShimmer} calculates the average absolute difference between the amplitude of the consecutive periods in a signal divided by the average amplitude. \emph{LocalDBShimmer}  is the average of the absolute value of 10-based logarithm of the difference between the amplitudes of consecutive periods in the signal, multiplied by 20. \emph{Apq3Shimmer} is the three-point Amplitude Perturbation Quotient: the average absolute difference between the amplitude of a period and the average of the amplitudes of its two neighbours -- left and right -- divided by the average amplitude. \emph{Apq5Shimmer} and \emph{Apq11Shimmer} are similar to \emph{Apq3Shimmer}, but uses data from four and 10 neighbours respectively instead of two. \emph{DdpShimmer} is three times the value of \emph{Apq3Shimmer} 
    
\subsubsection{MFCC:}
Mel Frequency Cepstral Coefficients (MFCC)~\cite{pols1977spectral} are used to understand the rate of energy changes in different spectrum bands of the speech: If a cepstral coefficient has negative value, it indicates that majority of spectral energy in that spectrum band is concentrated in the high frequencies; if a cepstral has positive value, it indicates that majority of spectral energy is concentrated in low frequencies. As we get several entries for each of the 13 spectral regions of MFCC, we take the mean (\textbf{MeanMFCC-[0-12]}) and mean variation (\textbf{VariationMFCC[0-12]}) for each of these spectral regions. 
% It is widely used in speech recognition, genre classification in music, similarity in audios measurement, etc.
\subsubsection{Relative Band Power:}
Relative band power features were calculated by checking how much power is present in four different spectrum of frequency windows in the range [0,500,1000,2000,4000] Hz. The power contained in these four regions are denoted by \textbf{RelBandPower[0-3]}.  Through applying FFT, we convert the audio signal into frequency domain. Then, we calculate the power contained in the frequencies belonging to each bucket, aggregate them and  calculate the median in each of these buckets 
% \W{Need to make sure that  it is correct from Dr Little}
\subsubsection{Harmonic-to-noise(HNR) ratio}
 HNR denotes the ratio of desired signal and background noise; higher HNR indicates better quality of audio.
 
\subsubsection{Recurrence period density entropy (RPDE):}
A perfectly recurrent time signal will maintain a strict time period. Recurrence period density entropy (RPDE)  determines how much a signal is maintaining a strict periodicity after  the signal is reconstructed in phase space~\cite{little2007exploiting}. By aggregating the time-periods recorded in our signal, and calculating the entropy of those time-periods, we get a measure of how much variation is present in those time-periods. A perfectly healthy voice will be able to maintain sustained vibration, hence it should have an entropy close to zero. Finally, the RPDE values are normalized in the range [0,1] to be used as feature.
% We can calculate the feature $H_{norm}$ through normalizing the RPDE values in the range [0,1].
\subsubsection{Detrended fluctuation analysis (DFA):}
As human voice is produced by turbulent air-flows through our vocal folds, degeneration of voice-fold structure (due to age or diseases) can produce increased noise in speech~\cite{little2007exploiting}.  
Detrended fluctuation analysis (DFA) measures the extent of the stochastic self-similarity of the noise in the speech signal produced due to possible alteration in vocal fold structure. These kind of noises can be represented through a statistical scaling component on a range of physical scales; this scaling component is comparatively larger for people with voice disorders~\cite{little2007exploiting,little2008suitability}. 

% The DFA algorithm first computes the amplitude variation F(L) over a range of rtime sclaes L; then the self-similarity of the speech signal is quantified through the slot of log-log plot of F(L) vs L. These slop values are normalized to the range [0,1] through a nonlinear transformation, producing the $alpha_{norm}$ feature.
\subsubsection{Pitch Period Entropy (PPE):}
~\cite{little2008suitability} introduces a new feature Pitch Period Entropy (PPE) to calculate the entropy present in the pitch of an audio signal. First, a standard time-signal of pitch is converted into the logarithmic domain to capture the logarithmic nature of speech generation and perception. Then, to remove the gender and person specific trends present in the pitch --  as we know that females have higher pitch voices than males, and there exists individual differences in pitch -- we apply a standard whitening filter. Then, we use calculate the probability density of the residual signal. For a healthy voice signal, most of the probabilities will be concentrated on a narrow range. However, the people with vocal disorders cannot maintain a sustained pitch for a long time, therefore, there probability distribution will be much more dispersed. This dispersion is calculated through entropy, which precisely calculates how much disorganization there is in a system. A lower entropy means that the pitch was sustained over a long time, a higher entropy indicates problems with the vocal cords and probably dysphonia as well.

% \MSI{Feature descriptions from Section 4.3 can be moved to supplementary material.}

\subsection{Embedding-features Extraction}
\label{ssec:pase_feats}
We extracted Problem Agnostic Speech Encoder (PASE) embeddings~\cite{Pascual2019} for our audio files. PASE represents the information contained in a raw audio instance through a list of encoded vectors. To make sure that the encoded vectors contain the same information as the input audio file, they decode various properties of the audio file which include, the audio waveform, the Log Power Spectrum, Mel-frequency cepstral coefficients (MFCC), four prosody features (interpolated logarithm of the fundamental frequency, voiced/unvoiced probability, zero-crossing rate, and energy), Local Info max, etc. from the encoded vectors. To decode all these properties successfully, the encoded vectors must retain the relevant information about the input audio file. As these properties represent the inherent characteristics of the input audio file rather than any task-specific features, they can be easily used to solve a host of downstream tasks like speech classification, speaker recognition, emotion recognition and as we demonstrate, PD detection.

% \subsubsection{PD relvenat features}
% Following the work in \W{Citation based on Max's reply}, we extracted these 39 features:
% \begin{enumerate}
%     \item f0m: median frequency, a.k.a. median fundamental frequency or pitch
%     \item f0j: jitter (mean variation of f0)
%     \item f0jr is the jitter using the median variation of f0
%     \item ash: Shimmer using the mean variation of amplitude
%     \item ashr: Shimmer using the median variation of amplitude
%     \item CEPM: mean Mel Frequency Cepstral Coefficients (13 coefficients named cepm0, cepm1, ..., cepm12 in the excel file)
%     \item CEPJ: mean variation of Mel Frequency Cepstral Coefficients (13 coefficients named cepj0, cepj1, ..., cepj12 in the excel file)
%     \item Hnorm: RPDE feature
%     \item alpha: DFA feature
%     \item ppe: PPE feature proposed by Dr. Little
%     \item relbandpower: Relative Band Power captures four spectral features--delta, theta, alpha, beta (4 elements named relbandpower0, relbandpower1, relbandpower2, relbandpower3 in the excel file)
%     \item f0std: the standard deviation of fundamental frequency f0
% \end{enumerate}

\subsection{Experiments}

For each of the feature sets, we applied a standard set of machine learning algorithms like Support-vector-machine (SVM)~\cite{cortes1995support}, XGBoost ~\cite{chen2016xgboost}, LightGBM~\cite{ke2017lightgbm}, and Random-Forest~\cite{ho1995random} Classifier to classify PD vs. non-PD. SVM separates out the data into several classes while maintaining a maximum possible margin among the classes. It can use the kernel trick to project the data on a more abstract hyper-plane and thus provide the model with more expressiveness. Random forest is built as an ensemble of Decision Trees; each decision tree builds a tree using a subset of features and learns if-else type decision rules to make a prediction. We also use XGBoost and LightGBM: two algorithms based on Gradient boosting where they build successively better models by refining the current models. eXtreme Gradient Boosting(XGBoost) provides a framework for fast, distributed gradient boosting while employing sophisticated heuristics for penalizing poorly performing trees and better use of regularization. LightGBM is a boosting algorithm  that employs leaf-wise tree growth and hence it can make a better estimation of the information gain through examining a smaller subset of data, and gain very equivalent accuracy very fast at the expense of lower memory usage.

We used a leave-one-out cross validation training strategy; using this strategy one sample of the dataset is left out and the other n-1 samples are used to create a model and predict the remaining sample. We used metrics like Binary Accuracy and AUC to report our model's performance. Area-Under-Curve (AUC) is the area under the ROC (Receiver Operating Characteristics) curve. The ROC curve is constructed by calculating the area under the curve produced by taking the ratio of the True-Positive-Rate and the False-Positive-Rate while varying the threshold of the decision. AUC can have highest value of 1, which denotes that the two classes can be separated perfectly; whereas an AUC value of 0.5 indicates that the model has no capability to distinguish between the two classes\footnote{https://towardsdatascience.com/understanding-auc-roc-curve-68b2303cc9c5} Since our dataset is imbalanced, AUC is a much better metric to understand the true performance of our model.

Since there exists significant imbalance between our PD and non-PD class and PD is the minority class, we have much more samples of non-PD than PD. Therefore, the model has a tendency to choose the majority non-PD class as a default which can yield a high False-Negative score, which is particularly bad in our case since our system is envisioned to be screening tool to help people get a preliminary screening for PD so that they can visit a doctor immediately. It will be quite problematic if we predict a PD patient as non-PD and thus provide him/her with a false sense of security and create a situation where his/her disease progress due to lack of medical care.
 To tackle that challenge, we used the Synthetic Minority Oversampling Technique (SMOTE)~\cite{chawla2002smote} and SVMSmote~\cite{nguyen2011borderline}. SMOTE can create synthetic data instances for the under-represented class. For each sample in the minority class, it calculates the K-nearest neighbours of that instance. Then, it calculates the straight line connecting each sample to all of its neighbours and then samples new synthetic data samples situated over that connecting line. SVMSMOTE  focuses on the boundary of different classes. We know that SVM creates a decision boundary around classes by trying to maintain the maximum possible margin among classes. Thus SVMSMOTE samples data in a manner than focuses on the boundary region of the minority class and samples data from that region such that the boundary between the classes is either expanded or consolidated.

\subsection{Model Interpretation}
\label{ssec:methods_model_interpretation}

 For interpreting the models, we are using the SHAP technique based on Shapley Value. Shapley value is a game-theoretic concept of distributing the payout fairly among the players~\cite{shapley1953value}. In the machine learning context, each individual feature of an instance can be thought of as a player and the payout is the difference between an instance's prediction and average prediction. It is based on rigorous mathematical foundation and it ensures fair distribution of importance amongst the features through ensuring the necessary mathematical properties. In principle, Shapley value can be computed by calculating the average marginal contribution for each feature across all the examples ~\cite{vstrumbelj2014explaining}.  For each feature, all possible combinations of all the other features -- defined as coalition -- is generated. Then for each sample generated from those coalitions, two kinds of scores are calculated: one where the intended feature is positive and another where it is not. The weighted average of the difference between these two scores is the Shapley Value denoting the contribution of that feature towards prediction.
 
However, computing Shapley value in this manner will take exponential time. Therefore, Lundberg et al 2017~\cite{lundberg2017unified} proposed SHAP(SHapley Additive exPlanations) based on Shapley Value with several additions. They added two methods called KernelSHAP that estimates Shapley values through kernel-based estimation and TreeSHAP that provides an exact, efficent estimation method for tree-based models like Decision Tree, Random Forest, etc. SHAP unifies the idea of Local  interpretable  model-agnostic  explanations  (LIME) and Shapley value by enabling us to build both local and global models: designed for explaining a portion of data samples and all data samples respectively. To utilize SHAP to explain gradient boosting and tree-based models like XGBoost, Lundberg et al. 2020~\cite{lundberg2020local} introduced methods to provide a polynomial time model to compute optimized explanations. Their method can generate local interpretations -- how the features impact one particular prediction of a data instance -- and then combine those local interpretations to make global interpretations about features present in the entire dataset. By setting up a class of features to condition on, they traverse the tree in the following manner: if we are traversing on a node that was split based on a feature we are conditioning on, we simply follow the decision path; otherwise, the results from the left and right sub-trees originating from the current node is computed recursively and their results added through a weighted summation strategy -- thus computing the SHAP value for the feature in question.

They validated their approach on three classical medical dataset -- modeling mortality risk from 20 years of follow up through  National Health and Nutrition Examination Survey (NHANES) I Epidemiologic Followup Study, classifying whether kidney patients will develop acute end-state renal diseases with 4 year through 
Chronic Renal Insufficiency Cohort (CRIC) study and predicting duration of hospital stay of a patient after an upcoming procedure.
Their models achieved competitive baseline performance while their explanations provided interpretable insights about the model's decision making process that is grounded in previous research. 

As SHAP constructs individual local interpretation and then provides an aggregated view of the results, we want to validate whether the features captured by SHAP are truly impactful. In order to do that, we took the features depicted in Fig ~\ref{fig:bar_three_models}.A in terms of their ranking and incrementally added them one by one in the feature set. For each feature set, we trained a new model and reported the performance in terms of AUC and Accuracy. As depicted in Fig.~\ref{fig:validating_shap_itself} through a three-point moving average trend-line, the value of AUC plateaued after we added 10 features. Thus, we can show that, SHAP can recognize the most features driving the model's performance.

\section{Data and Code Availability}
Please contact the corresponding author for getting access to the data and code. The maintenance and sharing of data collected through the PARK system must adhere to the Health Insurance Portability and Accountability Act (HIPPA) regulations. Therefore, it can only be shared by adding the interested party in the protocol maintained by the relevant Institutional Review Board (IRB).

\section{Competing Interests}
The authors declare that there are no competing interests.

\section{Author Contribution}
Wasifur Rahman, Sangwu Lee, Md Saiful Islam and Victor Nikhil Antony worked in data analysis, feature extraction, model training, model interpretation and manuscript preparation. Harshil Ratnu, Abdullah Al Mamun, Ellen Wegner,  Stella Jensen-Roberts helped build, maintain, and co-ordinate the data collection procedure. Mohammad Rafayet Ali, Max Little, and Ray Dorsey helped improve the manuscript, suggested important experiments and provided access to critical resources like code and data. Ehsan Hoque was the PI of the project; he facilitated the entire project and helped to shape the narrative of the manuscript.
%conclusion will be removed
%\input{sections/Conclusion}
%\input{sections/Discussions}
%\input{sections/Conclusion}
% \section{Acknowledgments}
% \W{There are some more sectiosn to be added. Check:https://www.nature.com/articles/s41746-020-0302-y
% https://www.nature.com/nature/for-authors/initial-submission}

\bibliography{ms}

\begin{thebibliography}{46}
\providecommand{\natexlab}[1]{#1}
\providecommand{\url}[1]{\texttt{#1}}
\expandafter\ifx\csname urlstyle\endcsname\relax
  \providecommand{\doi}[1]{doi: #1}\else
  \providecommand{\doi}{doi: \begingroup \urlstyle{rm}\Url}\fi

\bibitem[Ali et~al.(2020)Ali, Hernandez, Dorsey, Hoque, and
  McDuff]{ali2020spatio}
Mohammad~Rafayet Ali, Javier Hernandez, E~Ray Dorsey, Ehsan Hoque, and Daniel
  McDuff.
\newblock Spatio-temporal attention and magnification for classification of
  parkinson's disease from videos collected via the internet.
\newblock In \emph{2020 15th IEEE International Conference on Automatic Face
  and Gesture Recognition (FG 2020)(FG)}, pages 53--60, 2020.

\bibitem[Bandini et~al.(2017)Bandini, Orlandi, Escalante, Giovannelli,
  Cincotta, Reyes-Garcia, Vanni, Zaccara, and Manfredi]{bandini2017analysis}
Andrea Bandini, Silvia Orlandi, Hugo~Jair Escalante, Fabio Giovannelli, Massimo
  Cincotta, Carlos~A Reyes-Garcia, Paola Vanni, Gaetano Zaccara, and Claudia
  Manfredi.
\newblock Analysis of facial expressions in parkinson's disease through
  video-based automatic methods.
\newblock \emph{Journal of neuroscience methods}, 281:\penalty0 7--20, 2017.

\bibitem[Bhattarai et~al.(2017)Bhattarai, Prasad, Alsadoon, Pham, and
  Elchouemi]{bhattarai2017experiments}
Kritagya Bhattarai, PWC Prasad, Abeer Alsadoon, L~Pham, and Amr Elchouemi.
\newblock Experiments on the mfcc application in speaker recognition using
  matlab.
\newblock In \emph{2017 Seventh International Conference on Information Science
  and Technology (ICIST)}, pages 32--37. IEEE, 2017.

\bibitem[Boersma and Weenink(2018)]{praat}
Paul Boersma and David Weenink.
\newblock {P}raat: doing phonetics by computer [{C}omputer program].
\newblock Version 6.0.37, retrieved 3 February 2018
  \url{http://www.praat.org/}, 2018.

\bibitem[Center(2019)]{pew2019mobile}
Pew~Research Center.
\newblock Mobile fact sheet.
\newblock \emph{Internet \& Technology.}, 2019.
\newblock URL \url{https://www.pewresearch.org/internet/fact-sheet/mobile/}.

\bibitem[Chawla et~al.(2002)Chawla, Bowyer, Hall, and
  Kegelmeyer]{chawla2002smote}
Nitesh~V Chawla, Kevin~W Bowyer, Lawrence~O Hall, and W~Philip Kegelmeyer.
\newblock Smote: synthetic minority over-sampling technique.
\newblock \emph{Journal of artificial intelligence research}, 16:\penalty0
  321--357, 2002.

\bibitem[Chen and Guestrin(2016)]{chen2016xgboost}
Tianqi Chen and Carlos Guestrin.
\newblock Xgboost: A scalable tree boosting system.
\newblock In \emph{Proceedings of the 22nd acm sigkdd international conference
  on knowledge discovery and data mining}, pages 785--794. ACM, 2016.

\bibitem[Cortes and Vapnik(1995)]{cortes1995support}
Corinna Cortes and Vladimir Vapnik.
\newblock Support-vector networks.
\newblock \emph{Machine learning}, 20\penalty0 (3):\penalty0 273--297, 1995.

\bibitem[Duffy(2019)]{duffy2019motor}
Joseph~R Duffy.
\newblock \emph{Motor Speech Disorders E-Book: Substrates, Differential
  Diagnosis, and Management}.
\newblock Elsevier Health Sciences, 2019.

\bibitem[Goetz et~al.(2008)Goetz, Tilley, Shaftman, Stebbins, Fahn,
  Martinez-Martin, Poewe, Sampaio, Stern, Dodel, et~al.]{goetz2008movement}
Christopher~G Goetz, Barbara~C Tilley, Stephanie~R Shaftman, Glenn~T Stebbins,
  Stanley Fahn, Pablo Martinez-Martin, Werner Poewe, Cristina Sampaio,
  Matthew~B Stern, Richard Dodel, et~al.
\newblock Movement disorder society-sponsored revision of the unified
  parkinson's disease rating scale (mds-updrs): scale presentation and
  clinimetric testing results.
\newblock \emph{Movement disorders: official journal of the Movement Disorder
  Society}, 23\penalty0 (15):\penalty0 2129--2170, 2008.

\bibitem[Goldberger and Goldberger(1991)]{goldberger1991course}
Arthur~Stanley Goldberger and Arthur Stanley~Goldberger Goldberger.
\newblock \emph{A course in econometrics}.
\newblock Harvard University Press, 1991.

\bibitem[Haaxma et~al.(2007)Haaxma, Bloem, Borm, Oyen, Leenders, Eshuis, Booij,
  Dluzen, and Horstink]{haaxma2007gender}
Charlotte~A Haaxma, Bastiaan~R Bloem, George~F Borm, Wim~JG Oyen, Klaus~L
  Leenders, Silvia Eshuis, Jan Booij, Dean~E Dluzen, and Martin~WIM Horstink.
\newblock Gender differences in parkinson’s disease.
\newblock \emph{Journal of Neurology, Neurosurgery \& Psychiatry}, 78\penalty0
  (8):\penalty0 819--824, 2007.

\bibitem[Ho et~al.(1998)Ho, Iansek, Marigliani, Bradshaw, and
  Gates]{ho1998speech}
Aileen~K Ho, Robert Iansek, Caterina Marigliani, John~L Bradshaw, and Sandra
  Gates.
\newblock Speech impairment in a large sample of patients with parkinson's
  disease.
\newblock \emph{Behavioural neurology}, 11\penalty0 (3):\penalty0 131--137,
  1998.

\bibitem[Ho(1995)]{ho1995random}
Tin~Kam Ho.
\newblock Random decision forests.
\newblock In \emph{Proceedings of 3rd international conference on document
  analysis and recognition}, volume~1, pages 278--282. IEEE, 1995.

\bibitem[Howlett(2014)]{howlett2014neurology}
William~P Howlett.
\newblock Neurology in africa.
\newblock \emph{Neurology}, 83\penalty0 (7):\penalty0 654--655, 2014.

\bibitem[Jadoul et~al.(2018)Jadoul, Thompson, and de~Boer]{parselmouth}
Yannick Jadoul, Bill Thompson, and Bart de~Boer.
\newblock Introducing {P}arselmouth: A {P}ython interface to {P}raat.
\newblock \emph{Journal of Phonetics}, 71:\penalty0 1--15, 2018.
\newblock \doi{https://doi.org/10.1016/j.wocn.2018.07.001}.

\bibitem[Ke et~al.(2017)Ke, Meng, Finley, Wang, Chen, Ma, Ye, and
  Liu]{ke2017lightgbm}
Guolin Ke, Qi~Meng, Thomas Finley, Taifeng Wang, Wei Chen, Weidong Ma, Qiwei
  Ye, and Tie-Yan Liu.
\newblock Lightgbm: A highly efficient gradient boosting decision tree.
\newblock In \emph{Advances in neural information processing systems}, pages
  3146--3154, 2017.

\bibitem[Khadilkar(2013)]{khadilkar2013neurology}
SV~Khadilkar.
\newblock Neurology in india.
\newblock \emph{Annals of Indian Academy of Neurology}, 16\penalty0
  (4):\penalty0 465, 2013.

\bibitem[Kinnunen et~al.(2007)Kinnunen, Chernenko, Tuononen, Fr{\"a}nti, and
  Li]{kinnunen2007voice}
Tomi Kinnunen, Evgenia Chernenko, Marko Tuononen, Pasi Fr{\"a}nti, and Haizhou
  Li.
\newblock Voice activity detection using mfcc features and support vector
  machine.
\newblock In \emph{Int. Conf. on Speech and Computer (SPECOM07), Moscow,
  Russia}, volume~2, pages 556--561, 2007.

\bibitem[Kubota et~al.(2016)Kubota, Chen, and Little]{kubota2016machine}
Ken~J Kubota, Jason~A Chen, and Max~A Little.
\newblock Machine learning for large-scale wearable sensor data in parkinson's
  disease: Concepts, promises, pitfalls, and futures.
\newblock \emph{Movement disorders}, 31\penalty0 (9):\penalty0 1314--1326,
  2016.

\bibitem[Langevin et~al.(2019)Langevin, Ali, Sen, Snyder, Myers, Dorsey, and
  Hoque]{langevin2019park}
Raina Langevin, Mohammad~Rafayet Ali, Taylan Sen, Christopher Snyder, Taylor
  Myers, E~Dorsey, and Mohammed~Ehsan Hoque.
\newblock The park framework for automated analysis of parkinson's disease
  characteristics.
\newblock \emph{Proceedings of the ACM on Interactive, Mobile, Wearable and
  Ubiquitous Technologies}, 3\penalty0 (2):\penalty0 54, 2019.

\bibitem[Little et~al.(2008)Little, McSharry, Hunter, Spielman, and
  Ramig]{little2008suitability}
Max Little, Patrick McSharry, Eric Hunter, Jennifer Spielman, and Lorraine
  Ramig.
\newblock Suitability of dysphonia measurements for telemonitoring of
  parkinson’s disease.
\newblock \emph{Nature Precedings}, pages 1--1, 2008.

\bibitem[Little and Badawy(2019)]{little2019causal}
Max~A Little and Reham Badawy.
\newblock Causal bootstrapping.
\newblock \emph{arXiv preprint arXiv:1910.09648}, 2019.

\bibitem[Little et~al.(2007)Little, McSharry, Roberts, Costello, and
  Moroz]{little2007exploiting}
Max~A Little, Patrick~E McSharry, Stephen~J Roberts, Declan~AE Costello, and
  Irene~M Moroz.
\newblock Exploiting nonlinear recurrence and fractal scaling properties for
  voice disorder detection.
\newblock \emph{Biomedical engineering online}, 6\penalty0 (1):\penalty0 23,
  2007.

\bibitem[Logemann et~al.(1978)Logemann, Fisher, Boshes, and
  Blonsky]{logemann1978frequency}
Jeri~A Logemann, Hilda~B Fisher, Benjamin Boshes, and E~Richard Blonsky.
\newblock Frequency and cooccurrence of vocal tract dysfunctions in the speech
  of a large sample of parkinson patients.
\newblock \emph{Journal of Speech and hearing Disorders}, 43\penalty0
  (1):\penalty0 47--57, 1978.

\bibitem[Lonini et~al.(2018)Lonini, Dai, Shawen, Simuni, Poon, Shimanovich,
  Daeschler, Ghaffari, Rogers, and Jayaraman]{Lonini2018}
Luca Lonini, Andrew Dai, Nicholas Shawen, Tanya Simuni, Cynthia Poon, Leo
  Shimanovich, Margaret Daeschler, Roozbeh Ghaffari, John~A Rogers, and Arun
  Jayaraman.
\newblock {Wearable sensors for Parkinson's disease: which data are worth
  collecting for training symptom detection models}.
\newblock \emph{npj Digital Medicine}, 1:\penalty0 64, 2018.

\bibitem[Lundberg and Lee(2017)]{lundberg2017unified}
Scott~M Lundberg and Su-In Lee.
\newblock A unified approach to interpreting model predictions.
\newblock In \emph{Advances in Neural Information Processing Systems}, pages
  4765--4774, 2017.

\bibitem[Lundberg et~al.(2020)Lundberg, Erion, Chen, DeGrave, Prutkin, Nair,
  Katz, Himmelfarb, Bansal, and Lee]{lundberg2020local}
Scott~M Lundberg, Gabriel Erion, Hugh Chen, Alex DeGrave, Jordan~M Prutkin,
  Bala Nair, Ronit Katz, Jonathan Himmelfarb, Nisha Bansal, and Su-In Lee.
\newblock From local explanations to global understanding with explainable ai
  for trees.
\newblock \emph{Nature machine intelligence}, 2\penalty0 (1):\penalty0
  2522--5839, 2020.

\bibitem[M{\"u}ller(2007)]{muller2007information}
Meinard M{\"u}ller.
\newblock \emph{Information retrieval for music and motion}, volume~2.
\newblock Springer, 2007.

\bibitem[Nguyen et~al.(2011)Nguyen, Cooper, and Kamei]{nguyen2011borderline}
Hien~M Nguyen, Eric~W Cooper, and Katsuari Kamei.
\newblock Borderline over-sampling for imbalanced data classification.
\newblock \emph{International Journal of Knowledge Engineering and Soft Data
  Paradigms}, 3\penalty0 (1):\penalty0 4--21, 2011.

\bibitem[None(2020{\natexlab{a}})]{earlypd}
None.
\newblock Early onset parkinson’s disease.
\newblock
  \url{https://www.apdaparkinson.org/what-is-parkinsons/early-onset-parkinsons-disease/},
  2020{\natexlab{a}}.
\newblock Accessed: 2020-08-26.

\bibitem[None(2020{\natexlab{b}})]{pdIncidenceStat}
None.
\newblock Parkinson's foundation/statistics.
\newblock \url{https://www.parkinson.org/Understanding-Parkinsons/Statistics},
  2020{\natexlab{b}}.
\newblock Accessed: 2020-09-28.

\bibitem[on~Rating Scales~for Parkinson's~Disease(2003)]{movement2003unified}
Movement Disorder Society Task~Force on~Rating Scales~for Parkinson's~Disease.
\newblock The unified parkinson's disease rating scale (updrs): status and
  recommendations.
\newblock \emph{Movement Disorders}, 18\penalty0 (7):\penalty0 738--750, 2003.

\bibitem[Pascual et~al.(2019)Pascual, Ravanelli, Serrà, Bonafonte, and
  Bengio]{Pascual2019}
Santiago Pascual, Mirco Ravanelli, Joan Serrà, Antonio Bonafonte, and Yoshua
  Bengio.
\newblock {Learning Problem-Agnostic Speech Representations from Multiple
  Self-Supervised Tasks}.
\newblock In \emph{Proc. of the Conf. of the Int. Speech Communication
  Association (INTERSPEECH)}, pages 161--165, 2019.
\newblock URL \url{http://dx.doi.org/10.21437/Interspeech.2019-2605}.

\bibitem[Pols et~al.(1977)]{pols1977spectral}
Louis~CW Pols et~al.
\newblock Spectral analysis and identification of dutch vowels in monosyllabic
  words.
\newblock \emph{AmsterdamAcademische Pers}, 1977.

\bibitem[Poorjam et~al.(2019)Poorjam, Kavalekalam, Shi, Raykov, Jensen, Little,
  and Christensen]{poorjam2019automatic}
Amir~Hossein Poorjam, Mathew~Shaji Kavalekalam, Liming Shi, Yordan~P Raykov,
  Jesper~Rindom Jensen, Max~A Little, and Mads~Gr{\ae}sb{\o}ll Christensen.
\newblock Automatic quality control and enhancement for voice-based remote
  parkinson's disease detection.
\newblock \emph{arXiv preprint arXiv:1905.11785}, 2019.

\bibitem[Shapley(1953)]{shapley1953value}
Lloyd~S Shapley.
\newblock A value for n-person games.
\newblock \emph{Contributions to the Theory of Games}, 2\penalty0
  (28):\penalty0 307--317, 1953.

\bibitem[{\v{S}}trumbelj and Kononenko(2014)]{vstrumbelj2014explaining}
Erik {\v{S}}trumbelj and Igor Kononenko.
\newblock Explaining prediction models and individual predictions with feature
  contributions.
\newblock \emph{Knowledge and information systems}, 41\penalty0 (3):\penalty0
  647--665, 2014.

\bibitem[Titze and Martin(1998)]{titze1998principles}
Ingo~R Titze and Daniel~W Martin.
\newblock Principles of voice production, 1998.

\bibitem[Tsanas et~al.(2009)Tsanas, Little, McSharry, and
  Ramig]{tsanas2009accurate}
Athanasios Tsanas, Max~A Little, Patrick~E McSharry, and Lorraine~O Ramig.
\newblock Accurate telemonitoring of parkinson's disease progression by
  noninvasive speech tests.
\newblock \emph{IEEE transactions on Biomedical Engineering}, 57\penalty0
  (4):\penalty0 884--893, 2009.

\bibitem[Tsanas et~al.(2011)Tsanas, Little, McSharry, and
  Ramig]{tsanas2011nonlinear}
Athanasios Tsanas, Max~A Little, Patrick~E McSharry, and Lorraine~O Ramig.
\newblock Nonlinear speech analysis algorithms mapped to a standard metric
  achieve clinically useful quantification of average parkinson's disease
  symptom severity.
\newblock \emph{Journal of the royal society interface}, 8\penalty0
  (59):\penalty0 842--855, 2011.

\bibitem[Tsanas et~al.(2012{\natexlab{a}})Tsanas, Little, McSharry, and
  Ramig]{tsanas2012using}
Athanasios Tsanas, Max~A Little, Patrick~E McSharry, and Lorraine~O Ramig.
\newblock Using the cellular mobile telephone network to remotely monitor
  parkinsons disease symptom severity.
\newblock \emph{IEEE Transactions on Biomedical Engineering}, 9,
  2012{\natexlab{a}}.

\bibitem[Tsanas et~al.(2012{\natexlab{b}})Tsanas, Little, McSharry, Spielman,
  and Ramig]{tsanas2012novel}
Athanasios Tsanas, Max~A Little, Patrick~E McSharry, Jennifer Spielman, and
  Lorraine~O Ramig.
\newblock Novel speech signal processing algorithms for high-accuracy
  classification of parkinson's disease.
\newblock \emph{IEEE transactions on biomedical engineering}, 59\penalty0
  (5):\penalty0 1264--1271, 2012{\natexlab{b}}.

\bibitem[Van Den~Eeden et~al.(2003)Van Den~Eeden, Tanner, Bernstein, Fross,
  Leimpeter, Bloch, and Nelson]{van2003incidence}
Stephen~K Van Den~Eeden, Caroline~M Tanner, Allan~L Bernstein, Robin~D Fross,
  Amethyst Leimpeter, Daniel~A Bloch, and Lorene~M Nelson.
\newblock Incidence of parkinson’s disease: variation by age, gender, and
  race/ethnicity.
\newblock \emph{American journal of epidemiology}, 157\penalty0 (11):\penalty0
  1015--1022, 2003.

\bibitem[Yuan and Liberman(2008)]{yuan2008speaker}
Jiahong Yuan and Mark Liberman.
\newblock Speaker identification on the scotus corpus.
\newblock \emph{Journal of the Acoustical Society of America}, 123\penalty0
  (5):\penalty0 3878, 2008.

\bibitem[Yue et~al.(2020)Yue, Yang, Wang, Rahul, and Katabi]{Yuebodycompass}
Shichao Yue, Yuzhe Yang, Hao Wang, Hariharan Rahul, and Dina Katabi.
\newblock Bodycompass: Monitoring sleep posture with wireless signals.
\newblock \emph{Proc. ACM Interact. Mob. Wearable Ubiquitous Technol.},
  4\penalty0 (2), June 2020.
\newblock \doi{10.1145/3397311}.
\newblock URL \url{https://doi.org/10.1145/3397311}.

\end{thebibliography}
\bibliographystyle{plainnat}

\end{document}